# VORTEX DYNAMICS IN BULK HTS WITH LEVITATION TECHNIQUES

A. A. KORDYUK, V.V. NEMOSHKALENKO,
A. I. PLYUSHCHAY, R. V. VIZNICHENKO

Institute of Metal Physics
36 Vernadsky str, Kyiv 252680, Ukraine

## 1. Introduction

The opportunity of using superconductivity at liquid nitrogen temperatures that appeared after discovering high temperature superconductors (HTS) induced a number of investigations of different systems with levitation. Many interesting results about macroscopic magnetic properties of HTS were obtained but a real interest in these systems for large scale applications appears only with development of the melt-textured (MT) technology. The use of MT HTS in large scale systems such as flywheels for energy storage, electric motors and generators, permanent magnets, etc. is the most promising HTS application now. Such large grain HTS samples and levitation systems with them are actively studied now. But despite these efforts the nature of macroscopic magnetic behavior of such HTS samples is far from full understanding.

One reason for this is the complexity of a field configuration in large scale systems that makes it very difficult to fit the existing simple-geometry experiments and theoretic models to the real cases. The other one is the complexity of HTS bulks themselves. First of all, they have a granular structure. The flux trapping ability of HTS bulks, which generally forms a basis for their applications, depends fundamentally on the critical current density of the material and the length scale over which it flows. The grain boundaries and intragrain domains can complicate significantly the magnetic properties of these samples but even the quasi single crystalline HTS remain to be too complex for modeling the magnetic flux dynamics in them. The anisotropic structure of the HTS materials and variety of different pinning centers are the reasons for this.

Now, there are several groups in Europe [1-3], United States [4] and Japan [5] that succeeded in manufacturing of quasi single crystalline HTS bulks with a very strong pinning. They have obtained the critical current density more than $10^5$ A/cm$^2$ at liquid nitrogen temperatures. But there is a gap in study of magnetic properties of these materials. The properties which are under much consideration are the levitation force and the trapped magnetic field. These characteristics are very important for testing the samples quality, to obtain information about granularity and for such HTS applications as permanent magnets and levitators but they are not sufficient for complete describing the magnetic flux dynamics in these samples. This dynamics and its main characteristics, AC losses, are very important for both fundamental investigations of HTS and applications like flywheels and motors. However, studying these relatively large samples by conventional methods is a problem, since they are not only destructive but also apply only to small samples.

Levitation is not only an effective demonstration of superconductivity but also has a great potential to apply in science and technology, especially in an instrument construction branch. Besides their non-destructive nature, the systems with levitation are able to increase an accuracy of physical measurements; for example, the attempt to measure the quark charge using the mechanical oscillations of the superconducting sphere levitated in the magnetic field was described in [6].

During the last 10 years we were studying the macroscopic magnetic dynamics in HTS bulks by the different experimental techniques with levitation. We have developed a set of such techniques to investigate the properties of both granular HTS samples and melt-textured ones. There are three main techniques we use: (i) the resonance oscillations technique [7-18]; (ii) the high speed magnetic rotor [19-21]; (iii) the method of viscous motion of a permanent magnet through a HTS sample aperture [22,23]. The non-destructive nature of the first two methods gives us an opportunity to carry out the different experiments with the same samples and obtain interesting information about AC losses and vortex dynamics in granular HTS in a wide range of frequencies and AC field amplitudes. In this paper we give a short review of these techniques and the results obtained.

## 2. Techniques description

### 2.1. RESONANCE OSCILLATION TECHNIQUE

The resonance oscillations technique is simple to realize but allows to obtain the unique results in the investigations of magnetic properties of bulk HTS. This method is based on the investigation of a mechanical system with a freely levitated permanent magnet above an explored HTS sample. The basic system was used to investigate the granular [8-14] and composite [12] YBCO and BSCCO ceramics. To investigate the melt-textured YBCO samples the slight modification of it was used [15-18]. Scheme of such a basic system is represented in Fig.1.

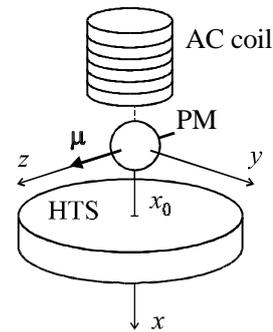

*Figure 1.* Scheme of the resonance oscillations technique.

Forced oscillations of the permanent magnet (PM) with mass $m$, magnetic moment $\mu$, and diameter $d$, which levitates on the distance of $x_0$ over the surface of the HTS sample were induced by the AC coil and their amplitude was measured by the microscope with the 5 $\mu$m accuracy. To increase the accuracy, the magnetic induction system was used. The PM produces the AC magnetic field $\mathbf{H} = \mathbf{H}_0 + \mathbf{h}_0 \sin(\omega t)$ at the surface and in the volume of the HTS. For the basic system we used: $m = 0.021$g, $\mu = 1.2$ G cm$^3$, $d = 1.2$ mm, $x_0 = 2.3$ mm.

The parameters of the magnetic flux dynamics in a HTS sample can be obtained from the dependencies of resonance frequency $\omega_0$ and dumping coefficient $\delta$ ($2\delta$ is the resonance curve width on a height of $A_{max}/\sqrt{2}$) on the resonance amplitude $A_{max}$ [6]. In the systems where it was possible to approximate a PM as a point magnetic dipole there are five modes $s$ of PM oscillations: three translation modes, $s = x, y, z$ (along the corresponding axes); and two rotation (or torsion) modes, $s = \psi$ and $\theta$ (around $x$ and $y$ axes respectively) [10-14]. Every mode has the following parameters of its own: resonance frequency $\omega$, damping $\delta$, and their dependencies on the PM amplitude $A$. For $Q$-factor, storage energy $W_0$ and energy loss per period $W$ we can write:

$$Q = \omega_0/2\delta, \ W_0 = 1/2 m\omega_0^2 A^2, \ W = 2\pi W_0/Q. \qquad (1)$$

All these parameters give us information about the magnetic flux dynamics in the HTS sample.

## 2.2. MAGNETIC ROTOR

The investigations of energy loss dependencies on AC magnetic field amplitude in HTS is important to understand the mechanism of the magnetic flux motion in such materials but the resonance oscillations technique gives information in a short frequency range only which depends on system geometry. The high speed rotor method allows to measure energy losses in a wide frequency range [19-21]. This method is based on a magnetic rotor on contactless HTS bearings [19]. Due to low energy consumption, this method allows to obtain information in the frequency range of 30-3000 Hz [20]. Fig. 2 represents the horizontal configuration of such an experiment.

In our experiments we used either the free spin down measurements [21, 24, 25] or phase difference detecting [19-21]. The first technique was used by Hull et al. [24-26] in the flywheel investigations to determine the coefficient of friction (COF). For the energy loss here we can write [21]:

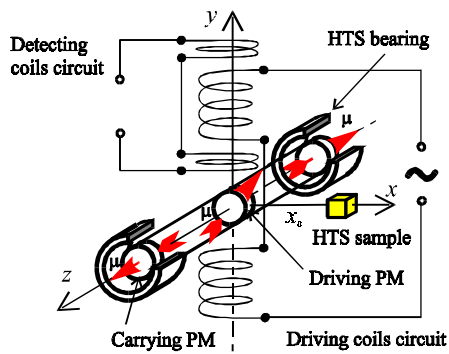

*Figure 2*. Horizontal configuration of the rotor for energy loss measuring.

$$W(\omega) = -2\pi J \dot\omega(\omega), \qquad (2)$$

where $J$ is the inertial moment of the rotor, $\omega$ is its rotation frequency, and $\dot\omega = d\omega/dt$ can be readily obtained from the experimental $\omega(t)$ dependency.

Another method of energy loss determination is the measuring of the phase $\varphi_0$ of the rotor rotation, which can be measured by phase difference between the driving and detecting coils (see Fig. 2). In this case the AC loss is [20, 21]

$$W(\omega) = \pi \mu H_0 \sin \varphi_0(\omega), \qquad (3)$$

where $\mu$ is the magnetic moment of the central driving magnet and $H_0$ is the AC field amplitude produced by the driving coils on this magnet.

Apart from these, there are different dynamic parameters of the rotor such as frequency and damping of the phase $\varphi_0$ oscillations, etc. that are determined by energy loss in the system and can be used for its measuring [21].

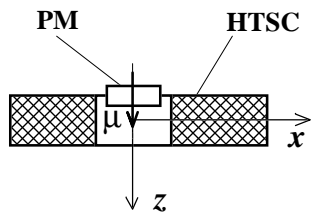

*Figure 3*. Scheme of the viscous motion effect.

### 2.3. VISCOUS MOTION METHOD

To investigate the magnetic flux relaxation with characteristic time about 1 sec in granular HTS the effect of viscous motion of a permanent magnet through the HTS aperture was used [22, 23]. Based on this effect the method for the investigation of a short time magnetic relaxation in granular HTS was developed. Scheme of the experiment presented in Fig. 3.

The magnetic moment of the PM was constant during all time. Its mass was increased by adding non-magnetic material. At 77 K under some relation of the geometry, PM mass and magnetic moment, the viscous motion of the PM through the aperture of the superconducting sample under gravity at 77 K takes place. Fig. 4 represents the PM velocity dependencies on time $v(t)$. The plateau, $v(t) = $ const, allows us to relate this velocity with characteristic time of the magnetic flux diffusion in HTS grains. In [22, 23] we obtained

$$v = \frac{g}{\omega_z^2 \tau_0}\left(1 - \frac{m_0}{m}\right), \qquad (4)$$

where $m$ is the PM mass; $m_0$ is a maximum mass at which the magnet motion velocity is zero that was obtained from the data extrapolation for the PM with the different masses; $g$ is the gravity acceleration; $\omega_z^2$ is the resonance frequency of the PM vertical oscillations. The method allows to measure the characteristic time of the magnetic flux relaxation with the accuracy up to 5% [23].

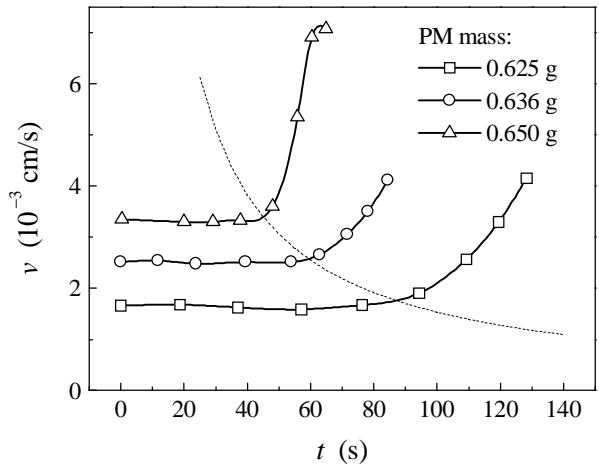

*Figure 4*. Experimental dependencies of the PM velocity on time for different PM masses for BSCCO granular sample at 77 K. Dashed line shows the region where the PM begins "to feel" the HTS bottom edge.

## 3. Granular HTS

By the resonance oscillations technique we investigated the ceramic $YBa_2Cu_3O_{7-x}$ superconductor with typical dimensions of component grains about 10–20 µm and $(Pb_{0.16}Bi_{0.84})_2Sr_2Ca_2Cu_3O_y$ with 5–10 µm grains. Two types of the samples were used in our experiments. They were the ceramic samples and the composite ones. All samples were 0.8 cm thick pellets of 4 cm diameter. The first ones were cut from the sintered superconductors. The composite samples were formed from dispersed HTS grains in an insulating paraffin. The dispersed grains were obtained by grinding the sintered superconductors. The absence of electrical contacts between grains was verified by the resistance measurements.

### 3.1. ROLE OF GRAINS

In [13] we have shown that from point of view of the elastic properties of PM-HTS systems, the Y- and Bi-superconductors under nitrogen temperatures can be considered as a set of independent isolated grains. For the resonance frequencies for all five modes we had obtained

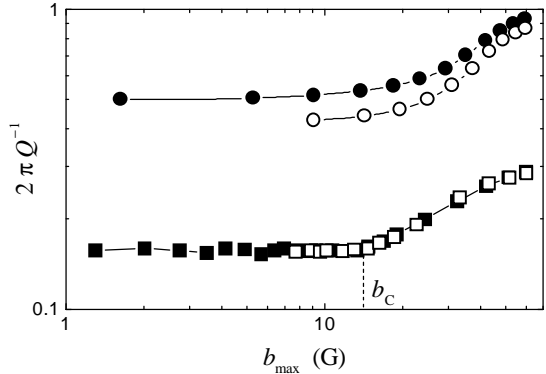

*Figure 5.* The inverse $Q$-factor *vs* the maximal amplitude of magnetic field on the HTS surface for the ceramic (closed symbols) and composite (open symbols) samples made of YBCO (squares) and BSCCO (circles) superconductors.

$$G_s \omega_s^2 = \alpha(\xi_s + \beta\zeta_s)\frac{\mu^2}{x_0^5}, \quad (5)$$

where $G_x = G_y = G_z = m$, $G_\psi = G_\theta = (1/10)\,m\,(d/x_0)^2$, $\xi_s$ and $\zeta_s$ are the geometrical coefficients represented in Table 1, $\alpha$ is the superconducting grains volume fraction, $\beta$ is a volume magnetization coefficient (see [13] for details). Thus, the coefficients $\alpha$ and $\beta$ can be determined from experimental values of resonance frequencies for any two modes.

In [12] we have shown that from point of view of AC losses, the Y- and Bi-superconductors also can be considered as a set of independent grains. Fig. 5 represents the inverse $Q$-factor as a function of $b_{\max}$, the maximum field amplitude on the HTS surface for the granular and composite samples. The $Q$-factor of this system is independent on $x_0$, $\alpha$ and $A$, so it is convenient to compare the different samples.

TABLE 1. Geometric coefficients of PM–HTS system with point dipole PM and granular HTS

| Modes $s$ | $\zeta_s$ | $\xi_s$ | $\xi_s^{\|}$ | $\xi_s^{\perp}$ | $W_{\|}/W$, % | $W_{\perp}/W$, % |
|---|---|---|---|---|---|---|
| $x$ | 3 / 16 | 3 / 16 | 193 / 1134 | 157 / 9072 | 91 | 9 |
| $y$ | −3 / 64 | 3 / 64 | 193 / 6804 | 2015 / 108864 | 61 | 39 |
| $z$ | −9 / 64 | 9 / 64 | 383 / 24300 | 48547 / 388800 | 11 | 89 |
| $\psi$ | −1 / 16 | 1 / 16 | 29 / 6480 | 47 / 810 | 7 | 93 |
| $\theta$ | −1 / 16 | 1 / 8 | 29 / 1620 | 347 / 3240 | 14 | 86 |

### 3.2. VISCOSITY EVALUATION

The plateau, that can be found on the dependencies of $Q^{-1}$ (Fig. 5) on the amplitude, shows the viscous mechanism of the energy losses for which $W = 2\pi m \delta \omega_0 A^2 \sim b_{\max}^2$ [10-12, 14]. Above the critical amplitude $b_c \sim 15$ G the hysteretic losses occur.

Outcoming from said, it is possible to evaluate the range of the viscosity of the intragrain flux motion. The linear viscosity $\eta_l$ is the proportionality factor between the viscous frictional force per vortex length and the vortex velocity: $\mathbf{f}_v = -\eta_l \mathbf{v}$. We also may write the volume viscosity $\eta_V = \eta_l B / \phi_0$, where $B$ is the magnetic field in the superconducting grains and $\phi_0$ is the magnetic flux quantum. Then $W = \pi \eta_l \omega \langle a^2 \rangle V_0 B / \phi_0$, where $\langle a^2 \rangle$ is the mean square oscillation amplitude of vortices and $V_0$ is the volume of the superconductor in which the main energy dissipation occurs. For a quantitative estimation of $\eta_l$ we have to know the mechanism of the vortex penetration into grain but we may estimate the lower limit of $\eta_l$. In [12] we have obtained $\eta_l > 3 \cdot 10^{-4}$ g/cm·s (or $\eta_V > 10^{-5}$ g/cm$^3$·s) that exceeds flux flow viscosity $\eta_{FF} \sim 10^{-7}$ g/cm·s [7] more then three orders of magnitude. Such giant viscosity may be explained by other mechanism of the viscous flux motion, namely TAFF [27, 28].

### 3.3. TRANSLATIONAL AND ROTATIONAL FIELD COMPONENTS

For the viscous flux motion the imaginary part of magnetic susceptibility $\chi''$ does not depend on AC field amplitude $h_0$ but only on its frequency. The energy loss per period can be obtained by taking integral over all HTS sample volume $V$:

$$W = \pi \alpha \chi'' \int_V h_0^2(\mathbf{r}) d\mathbf{r}^3. \qquad (6)$$

So, due to viscous character of the AC losses under $b_c$ in the PM-HTS system for each mode $s$ ($\mathbf{h}_s = (\partial \mathbf{H}/\partial s)A$) we can write [9, 11, 14]:

$$W_s = 4\pi^2 \alpha \mu^2 x_0^{-5} \xi_s \chi''(\omega_s) A^2 \qquad (7)$$

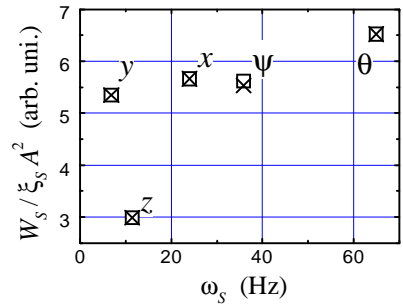

*Figure 6.* Experimental (□) and calculated (×) values of AC loss per square amplitude *vs* resonance frequency for five modes.

For the rotational modes $A = \varphi/x_0$, where $\varphi$ is the angular amplitude of the PM oscillations. The numeric values of $\xi_s$ are the same as in Eq. 5 and presented in Table 1. The dependence $\chi''(\omega_s) \propto W_s/A^2\xi_s$ shown in Fig. 6 is far from a real physical interpretation. So, as a first possible step, we have to separate the contributions of the translational $h_\parallel$ (parallel to DC field $\mathbf{h}_\parallel \| \mathbf{H}_0$) and rotational $h_\perp$ (perpendicular, $\mathbf{h}_\perp \perp \mathbf{H}_0$) AC field components. Then $h^\sim = h_\parallel^\sim + h_\perp^\sim$ and $W = W_\parallel + W_\perp$. The numeric values of $\xi_s^\parallel$ and $\xi_s^\wedge$ are also presented in

Table 1. The relative contribution from this components to AC loss is different from mode to mode (see Table 1). For the *x* and *y* modes the transverse component predominates; for the other tree modes, *z*, $\psi$ and $\theta$, it is conversely.

Now, from the experimental values of $W_s$ using numerical calculations we can find $\chi^2_\parallel(\omega)$ and $\chi^2_\wedge(\omega)$ functions. The result of such calculations for the YBCO sample is shown in Fig. 7. The calculations were made for the four modes and checked for the fifth one (see

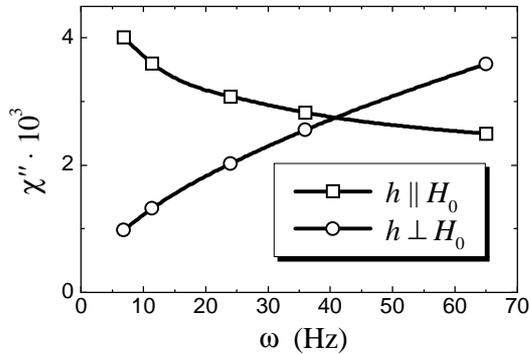

*Figure 7*. Imaginary part of susceptibility vs alternating magnetic field frequency for longitudinal and transverse field components for YBCO granular sample.

Fig. 6). These data were also verified in the direct measurements where the resonance frequency was changed by changing the PM mass [11, 14].

The differences in the frequency dependencies between the translational and rotational components are obvious if we remember that the translational component tends to change the magnitude of the intragrain magnetic flux by trying to penetrate through the grain boundary but the rotational component causes the change in the intragrain magnetic flux direction only (see Fig. 8). It means that the differences between $\chi^2_\parallel(\omega)$ and $\chi^2_\wedge(\omega)$ functions can be explained by the grain surface barrier for the magnetic flux to penetrate into grains [29-31].

### 3.4. VISCOUS ENERGY LOSSES

The results presented above show the viscous nature of the energy loss in granular HTS for the amplitudes less then 15 G for the vortices motion both in the grains volume and through the grains surface. It was shown that the energy losses are induced by TAFF with very high viscosity, but to determine this viscosity it is necessary to find a maximum on an energy loss frequency dependence.

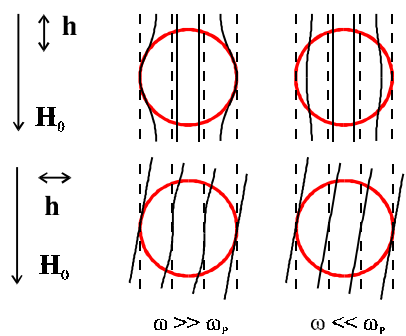

*Figure 8*. Penetration of parallel and perpendicular AC field components into superconducting grain.

To investigate the dependence of $W_\perp(\omega)$ the high speed magnetic rotor was used [20]. On the explored sample surface (Fig. 2), the rotor induces the superposition of the rotational and translational AC magnetic field components, but over ~ 100 Hz, the contribution of the translational component is negligible and it is possible to consider that all losses in HTS are induced by the viscous vortices motion in the grains volume.

The obtained results were published in [20] and are in good agreement with the critical state model with the large viscosity [32]. The obtained numeric value $\eta_l = 8 \cdot 10^{-5}$ g/cm s coincides with the above estimation [12]. The characteristic time of the rotational field component penetration is $10^{-3} - 10^{-2}$ s.

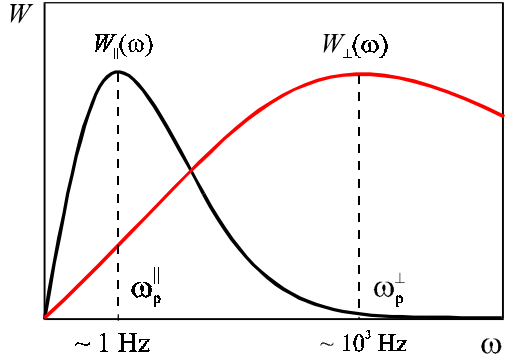

*Figure 9.* Schematic frequency dependencies of AC loss for parallel and perpendicular AC field components.

Additional information on the $W_\parallel(\omega)$ dependence can be obtained by the PM viscous motion method. Under velocities of about $10^{-3}$ cm/s the perpendicular component is negligible (Fig. 7, 9) and a slow viscous motion of the PM through the HTS aperture is mainly determined by the thermal activated diffusion of the magnetic flux into grains through the surface barrier [22]. From the characteristic time $\tau_0 = 1 \pm 0.07$ s obtained from (4) [22, 23] we can conclude that the viscosity of this process is by two-three orders of magnitude higher than that of in grains and $\eta_l \approx 10^{-2}$ g/cm s. The $W_\parallel(\omega)$ and $W_\perp(\omega)$ dependencies are schematically represented in Figure 9.

## 4. Melt-textured HTS

The melt-textured large grain HTS samples that are actively studied now are very different from the granular ones in the levitation properties. First, from point of view of different levitation systems they are very close to ideally hard superconductor — they have very strong pinning resulting in the absence of the effect of the PM rise above HTS sample at its cooling. Second, the small isolated grains approximation does not work for large grains [15].

### 4.1. IDEALLY HARD SUPERCONDUCTOR

In [15-18] we have shown that the approach of an absolutely hard superconductor is very useful to describe the elastic properties of levitation systems with the melt-textured HTS and as a first approximation to calculate the AC losses. The sense of this approach is to use the surface shielding currents to calculate the magnetic field distribution outside the superconductor and to obtain from this the elastic properties of the PM—HTS system. The magnetic field **B**(**r**) inside such an ideal superconductor does not change with any PM displacements. The feasibility of this approximation is roughly determined by the condition $d \ll L$, where $d$ is the field penetration depth and $L$ is the character system dimension (mainly the distance between PM and HTS). With such an approximation this problem has an exact analytical solution in case of a magnetic dipole over a flat superconductor in the field cooled (FC) case.

The nice illustration of this approach is the advanced mirror image method that was described in [15]. Its distinction from the usual one, which is applied to the type-I superconductors, is in using of the frozen PM image that creates the same magnetic field distribution outside the HTS as the frozen magnetic flux does. Fig. 10 illustrates this method. This configuration of the images shows that the ideally hard superconductor in FC case shields the magnetic field changes that induced by magnet displacements. The rigorous prove of this method is given in [16].

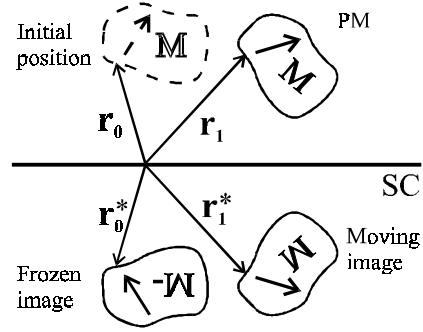

*Figure 10.* Advanced mirror image method.

But despite on so simple illustration the analytical solutions can be obtained for the interaction of the point magnetic dipole with a superconductor for some of the simplest geometry such as a plane or an ellipsoid only. In any other cases we have to use numerical calculations. Some analytical solutions for the levitation forces, resonance frequencies and non-linearity coefficients in the case of PM dipole over a flat superconductor were obtained in [15, 16].

The next step of the hard superconductor approximation is from zero-thickness of the undersurface shielding currents layer to the finite but thin enough (in respect to other system dimensions) layer. In this case the hysteresis of the elastic properties and AC losses can be obtained.

### 4.2. ENERGY LOSSES

For the melt-textured HTS where the energy losses have predominantly hysteretic nature [16, 17] the feasibility of such an approximation can be determined from the critical state model (the thickness of the layer $d$ carrying the critical current $J_c$ must be well less than the PM—HTS distance $z_0$ ) and for the above configuration we have

$$z_0 >> d = \frac{c}{4\pi} \frac{h(\boldsymbol{\rho}, s)}{J_c}, \qquad (8)$$

where $c$ is speed of light. As this takes place, and as the dimension of the HTS sample is much more than $d$, the energy loss per square $w(\boldsymbol{\rho}) = (c/24\pi^2) h^3(\boldsymbol{\rho})/J_c$. Then after integrating this energy loss over superconducting surface for $z$-mode (vertical PM oscillations)

$$W = \frac{2c}{3\pi} \frac{A^3}{J_c} \int_0^\infty \left(\frac{dH_r}{dz}\right)^3_{z=z_0} r\,dr, \qquad (9)$$

where $H_r$ is the perpendicular to $z$-axes PM field component. From here, the inverse $Q$-factor of the PM—HTS system

$$Q^{-1}(A) = \frac{2c}{3\pi^2} \frac{A}{J_c m \omega_0^2} \int_0^\infty \left(\frac{dH_r}{dz}\right)^3_{z=z_0} r\,dr. \qquad (10)$$

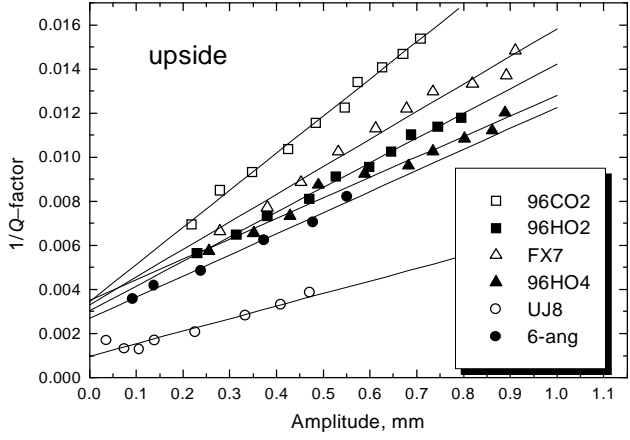

*Figure 11.* Experimental data of inverse *Q*-factor *vs* PM amplitude for upper sides of different MT HTS samples.

The condition (8) is much stronger than it is necessary to validate the use of the described approach for the melt-textured HTS. Even for $J_c \sim 10^4$ A/cm$^2$ and for $h \sim 100$ Oe the penetration depth $d \sim 0.1$ mm.

The experimental data obtained with the slightly modified resonance oscillation technique [17] are presented in Fig. 11. The linear dependencies for the inverse *Q*-factor ($Q^{-1} \sim \alpha + \beta A$) mean that energy losses have two components: $W = \alpha A^2 + \beta A^3$. The first component (viscous or surface losses [18]) needs the further investigations from the side of its dependence on the frequency. The second one is well known hysteretic losses that described by the critical state model. From this part and from Eq. (10) the value of the critical current density can be obtained.

But such linear dependencies of $Q^{-1}(A)$ as represented in Fig. 11 are observed only for the samples just after manufacturing. With time, due to the degradation of the undersurface layer, these dependencies are changed becoming non-linear [18]. It shows that the critical current density $J_c$ becomes spatially dependable. Fig. 12 presents such dependencies for one MT HTS sample just after manufacturing and after half a year. For the initial sample with the above approach we have obtained the value $J_c = 3.5 \cdot 10^4$ A/cm$^2$. We can also estimate the under-surface $J_c(x \to 0)$ of the degraded sample from the initial slope of its $Q^{-1}(A)$ dependence: $J_c(0) = 0.79 \cdot 10^4$ A/cm$^2$. Below we investigate a possibility to relate $J_c(x)$ and $Q^{-1}(A)$ functions.

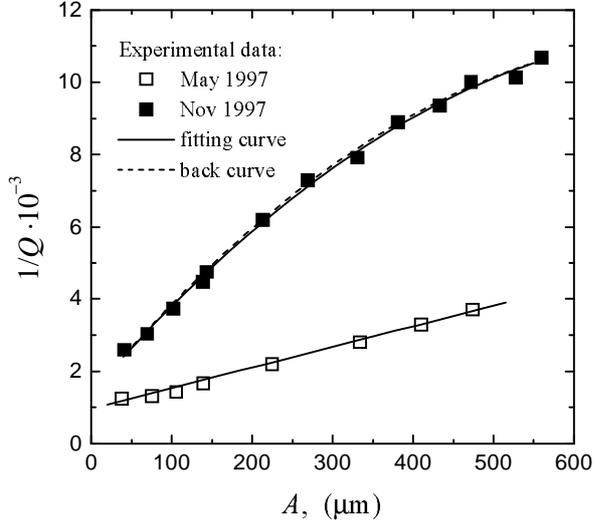

*Figure 11.* Experimental dependencies of the inverse *Q*-factor on the resonance amplitude of the permanent magnet above the bulk YBCO sample just after manufacturing (□) and in half a year (■).

## 4.3. RECONSTRUCTION OF CRITICAL CURRENT DENSITY PROFILES

In [18] we have developed the method of reconstruction of the critical current profiles from measurements of the hysteretic AC loss. We have shown that in the AC magnetic field the energy dissipated per period per unit of surface area in non-uniform superconductor ($J_c(B)$ = const, $J_c(x) \neq$ const) is

$$W_s(b_{0s}) = \frac{1}{\pi} \int_0^{x_p} b_0(x)\,(b_{0s} - b_0(x))\,dx. \qquad (11)$$

Here $x_p$ is the AC field penetration depth, $b_{0s}$ is the amplitude of the AC field at the superconducting surface, and the AC field amplitude inside a superconductor is

$$b_0(x) = b_{0s} - \frac{4p}{c} \int_0^x J_c(x)\,dx, \text{ for } 0 < x < x_p. \qquad (12)$$

Then, from (11) and (12) we can readily obtain the critical current density $J_c$ at the depth $x_p$ as a function of the experimental data $W_s(b_{0s})$:

$$J_c(x_p(b_{0s})) = \frac{c}{4\pi^2}\left(\frac{d^2 W_s}{db_{0s}^2}\right)^{-1} b_{0s}. \qquad (13)$$

Strictly speaking, Eq.(13) gives the dependence $J_c(b_{0s})$ at a depth $x_p(b_{0s})$ that provides some information about the real critical current profile but to obtain $J_c(x)$ itself in general case numerical calculations are needed.

It is necessary to mention that the similar approach for reconstruction of critical current density profiles from AC susceptibility measurements was proposed by Campbell in 1969 [33].

Let us determine $J_c(x)$ from the experimental data presented in Fig. 11 for the degraded sample that was obtained by the resonance oscillation technique with slightly modified geometry: a permanent magnet (SmCo$_5$, $\varnothing 6.3 \times 2.3$ mm, $m = 0.6$ g, $\mu = 38$ G cm$^3$) levitates above the melt-textured HTS sample ($\varnothing 32 \times 16$ mm) with the magnetic moment perpendicular to the surface. We use a field cooled case with the initial distance between the permanent magnet and the HTS surface $x_0 = 4.8$ mm. The system is symmetrical about the vertical $x$-axis. A more detailed description of this configuration was given in [17].

We will consider the hysteretic part only that can be fitted for this sample by the polynom $Q^{-1}(A) = q_1 A - q_2 A^2$, where $q_1 = 0.26$ cm$^{-1}$ and $q_2 = 1.65$ cm$^{-2}$.

The inverse $Q$-factor can be expressed as

$$Q^{-1}(A) = \frac{2}{m\omega^2 A^2} \int_0^{R_s} r\,W_s(b_{0s}(r,A))\,dr, \qquad (14)$$

where $\omega$ is the resonance frequency of the PM vertical oscillations, that slightly depends on $A$: $\omega = \omega_0 - \gamma A$, $\omega_0 = 36.3$ Hz, $\gamma = 0.1$ Hz/cm, and $R_s$ is the radius of the HTS sample. For a flat extremely hard superconductor (see [16]) at its surface we can write $b_{0s}(r,A) = 2\theta(r)A$, where $\theta(r) = dB_r(r)/dx$, and $B_r$ is the component of the permanent magnet field that is parallel to the surface of the sample. For simplicity we assume that $b_{0s} = 2\theta_{eff}A$ and use the value $S W_s$ instead of the integral in (14). Then

$$W_s(b_{0s}(A)) = \frac{\pi m \omega_0^2 b_{0s}^2}{4\theta_{eff}^2 S} Q^{-1}(b_{0s}(A)). \qquad (15)$$

We used the empirical formula $S = 2\pi k R_s x_0$, and in our configuration, $\theta_{eff} = 0.87\, \theta_{max}$ = 870 G/cm, $k = 0.94$. Then, using this assumption we obtain $J_c(x_p(b_{0s})) = J_{c0}/(1-\varepsilon b_{0s})$ with $\varepsilon = q_2/(\theta q_1)$ and finally [18]

$$J_c(x) = \frac{J_{c0}}{\sqrt{1 - p x}}, \qquad (16)$$

where $J_{c0} = c\theta_{eff}^3 S/(3\pi^3 m\omega_0^2 q_1) = 0.79 \cdot 10^4$ A/cm$^2$, and $p = 8\theta_{eff}^2 S q_2/(3\pi^2 m\omega_0^2 q_1^2) = 147$ cm$^{-1}$. The function $J_c(x)$ is shown in Fig. 12.

The dashed curve ("back" curve) in Fig. 11 represents the dependence obtained back from (16), (12) and (11) with real integrating in (14). The values $\theta_{eff}$ and $k$ were chosen for this curve to coincide with experimental data in two "points": for the initial slope determined by $J_{c0}$, and for $Q^{-1}(A_{max})$. The good coincidence of this back curve with the fitted curve on the hole range gives the proof of feasibility of Eq. (15) instead of (14).

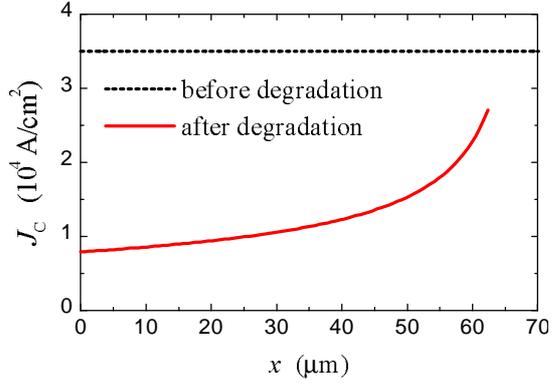

*Figure 12.* Critical current density profiles inside the HTS sample before (dashed line) and after (solid line) degradation calculated from AC loss experimental data.

## 5. Conclusions

In this paper we have presented a short review of the experimental techniques with levitation we have developed to investigate the macroscopic magnetic properties of the granular and quasi-monocrystalline HTS and some results of such an investigation. These methods are easy to use and allow to obtain the unique information about physical properties of HTS with the high accuracy, which sometimes is difficult to obtain by other methods.

The results obtained by these techniques can help to design different practical applications. For instance, we employed the results described above to create the compound non-contact bearing with self-centering and intrinsic damping of parasitic vibrations with both melt-textured and granular HTS components [21]. Scheme of this is represented in Fig. 13. The melt-textured components were used as a main carrying system to support the weight of the rotor and to provide the low AC losses at its rotations. The granular HTS were used to obtain the self-centering effect during cooling and to decrease parasitic resonances in the system. The AC losses in depends on the rotation speed of the rotor in the systems with and without the granular component are represented in Fig. 14 [21].

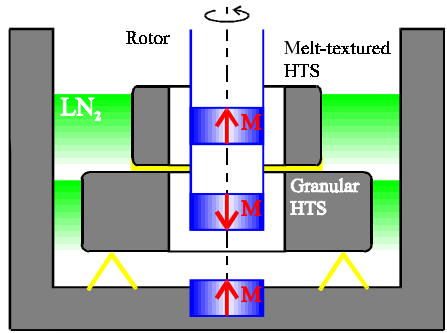 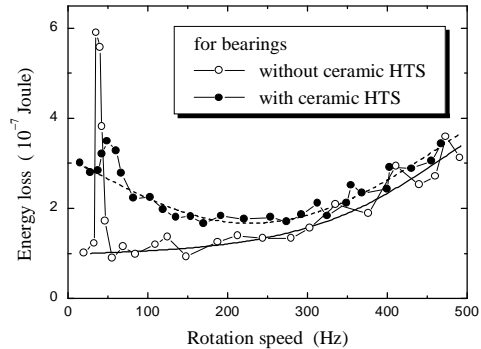

*Figure 13.* Scheme of the compound bearing.   *Figure 14.* Energy losses in the system with and without the granular component.

The differences we have found in the melt-textured samples prepared by the different groups relates generally to the surface and thin undersurface layer properties, but these differences are very important to understand the bulk HTS magnetic properties: energy losses, magnetic flux penetration, etc. These properties have the crucial significance for the different HTS large scale applications, but they are also very important from point of view of fundamental investigations.